\documentclass[12pt,twoside]{article}
\usepackage{fleqn,espcrc1}

\usepackage{graphicx}
\usepackage[figuresright]{rotating}

\def\be{\begin{eqnarray}}
\def\ee{\end{eqnarray}}
\def\bea{\begin{array}}
\def\eea{\end{array}}
\def\bei{\begin{itemize}}
\def\eei{\end{itemize}}
\newcommand{\AmS}{{\protect\the\textfont2
  A\kern-.1667em\lower.5ex\hbox{M}\kern-.125emS}}

\title{Chiral Symmetry and the Low-Energy Spectrum of the QCD Dirac Operator}

\author{J.J.M. Verbaarschot,\\
        Department of Physics and Astronomy,\\
        SUNY at Stony Brook, Stony Brook, NY 11794}
\begin{document}

% typeset front matter
\maketitle

\begin{abstract}

The order parameter of the chiral phase transition is directly related to 
the infrared part of the spectrum of the QCD Dirac operator. This part of 
the spectrum follows from the low energy limit of QCD which is given by 
a partition function of weakly interacting Goldstone modes. 
We find that the slope of the Dirac spectrum is determined by the
pion decay constant whereas for $\lambda \ll 1/L^2 \Lambda_{\rm QCD}$ 
the correlations of the Dirac eigenvalues are given by a random matrix 
theory with the global symmetries of the QCD partition function. 
A possible observation of these continuum 
results in lattice QCD with staggered fermions
is discussed.
\end{abstract}

\section{Introduction}
It has been well established that QCD is the correct theory of
the strong interactions. 
The vast body of evidence from lattice QCD simulations
at low and intermediate energies \cite{DeTar} is complemented by
perturbative calculations which become reliable at high energies where the
renormalized coupling constant is small due to asymptotic freedom. 
In this lecture we will focus on another limit where QCD simplifies. 
Because of spontaneous breaking
of chiral symmetry, QCD at low energy reduces to a theory of weakly
interacting Goldstone bosons. Although this theory cannot be derived from
QCD by means of an ab-initio calculation, its Lagrangian 
is determined uniquely by chiral symmetry and Lorentz invariance.
The validity of this low-energy theory is based on the presence of a mass-gap
which is a highly nontrivial and nonperturbative feature of QCD.

One of the questions we wish to 
address is to what extent the existence of an effective
low-energy theory imposes consistency conditions on the original theory.
This question was first posed in \cite{LS} for the
mass dependence of the QCD partition function. They argued that for
small quark masses it can be both obtained from the effective 
partition function and from QCD. Since the mass dependence of the QCD partition
function is given by the average of the fermion determinant this imposes
consistency conditions on the average properties of the
eigenvalues of the QCD Dirac operator.

Some of the properties of the Dirac eigenvalues are 
robust against {\it large} deformations of the gauge field action well
outside the scaling window.  
This type of spectral universality has been investigated 
primarily within the context
of Random Matrix Theory \cite{Hack,ADMN}. 
What has been found is that correlations
of eigenvalues on the scale of the average level spacing are
universal, i.e. they are robust against 
deformations of the probability distribution of the matrix elements.
The low-energy effective action is also insensitive to
{\it large} deformations of the gauge field action. 
The reason is the existence of a mass 
gap. In the next section we will relate this property 
to spectral universality. 

\section{Valence Quarks and the Low-Energy Limit of QCD} 

Because the same mass occurs both in the 
quark propagator and in the fermion determinant
the average Dirac spectrum cannot be obtained directly from the QCD
partition function.  In order to access the
spectrum of the Euclidean Dirac operator $D$
one has to introduce the valence quark mass dependence
of the chiral condensate defined by \cite{Vplb,Osbornprl,OTV}
\be
\Sigma(m_v) = \langle {\rm Tr} \frac 1 {m_v + D} \rangle.
\ee
The average is over the Euclidean QCD action 
which includes a fermion determinant
that depends on the the sea-quark masses. The spectral density 
per unit space-time volume $V= L^4$ of the Dirac operator $D$ 
is directly related to to $\Sigma(m_v)$,
\be
\rho(\lambda)/V = \frac 1{2\pi} \left (\Sigma(i\lambda +\epsilon) -
\Sigma(i\lambda -\epsilon) \right ).
\ee
The generating function for $\Sigma(m_v)$
is defined by \cite{pqChPT,OTV,DOTV}
\be
Z^{\rm pq}(m_v,J)  ~=~ \int\! [dA]
~\frac{\det(D +m_v+J)}{\det(D +m_v)}\prod_{f=1}^{N_{f}}
\det(D + m_f) ~e^{-S_{YM}[A]} ~.
\label{pqQCDpf}
\ee
In addition to the usual quarks this partition function contains
valence quarks and its bosonic superpartners. 
The chiral condensates for the different (super-)flavors 
are given by the same expression in terms of the eigenvalues
of the Dirac operator. We thus have a maximum breaking of 
the axial flavor symmetry.
The low-energy modes are then given by the Goldstone modes associated
with the spontaneous breaking of this  symmetry. 
Their quark content can be either
two sea-quarks, one sea quark and one valence quark, or two valence quarks.
The low-energy effective partition 
follows from the flavor super-symmetry and its breaking and
Lorentz invariance as is the case for the usual 
chiral Lagrangian \cite{pqChPT,OTV,DOTV}.
One major difference is that in this case the Goldstone manifold is 
a super-manifold with both compact and non-compact 
degrees of freedom \cite{Martin,OTV,DOTV}.

\section{Scales in the Dirac Spectrum}

For a nonzero value of the chiral condensate $\Sigma$ we can identify 
three important scales in the Dirac spectrum. The 
first scale is the smallest nonzero eigenvalue of the Dirac operator
given by $\lambda_{\rm min} =1/\rho(0)=\pi/\Sigma V$.
The second scale is the valence quark mass for which the Compton wavelength of 
the associated Goldstone bosons is 
equal to the size of the box. Using the Gell-Mann-Oakes-Renner relation
 we obtain as Thouless energy\cite{Altshuler,GL,Vplb,Osbornprl} 
\be
m_c = \frac {F^2}{\Sigma L^2},
\ee
where $F$ the pion decay constant.
A third scale is given by a typical hadronic mass scale. The three scales 
are ordered as $\lambda_{\rm min} \ll m_c \ll \Lambda$.
For valence quark masses $m_v \ll m_c$ the kinetic term in the effective 
action can be neglected and the low-energy partition function can be 
reduced to a zero dimensional integral. 
However, any partition function with a mass gap and
the same pattern of chiral symmetry breaking as in QCD can be reduced this way. 
The simplest such theory is chiral Random Matrix Theory (chRMT) 
\cite{SVR}. In that case spontaneous breaking of chiral
symmetry arises in the limit of infinite matrices.  
The advantage of working with chRMT is that is relatively simple to derive
the distribution of the smallest eigenvalues. 
Of course, the results for $\Sigma(m_v)$ 
obtained from the partially quenched partition function and from chRMT 
coincide \cite{OTV,DOTV}. 

The kinetic term is also determined uniquely by chiral symmetry and Lorentz
invariance which allows us to calculate the Dirac spectrum in the domain
$m_v \ll \Lambda$. This results in the slope of the Dirac spectrum at
$\lambda = 0$ which for $N_f$ massless flavors is given by \cite{OTV,TV}
\be
\frac{\rho'(0)}{\rho(0)} =
\frac {(N_f-2)(N_f+\beta)}
{16\pi \beta}\frac{\Sigma_0 }{F^4}.
\ee 
Here, $\beta$ denotes the Dyson index of the Dirac operator. For QCD with
fundamental fermions and three or more colors (with $\beta =2$) this result
was first derived in \cite{Smilstern}. The other two
possibilities, $\beta =1 $ and $\beta =4$, refer to QCD with fundamental
fermions and two colors and QCD with adjoint fermions and two or more colors,
respectively.

The domain below $m_c$ has been investigated extensively
by means of lattice QCD 
simulations and agreement with the chRMT results has been found
\cite{Vplb,Tilo,Hip,Heller,hiplat99,Tiloval,Poulval,Karlval,Damtopo}. 
A somewhat 
surprising result is that the lattice QCD data reproduce the analytical
result for zero topological charge. This will be explained 
in the next section.

\section{Approach to the Continuum Limit for Staggered Fermions}

The low-energy limit of QCD and the small Dirac
eigenvalues are described by the same partition function.
In order to recover the continuum $U_A(1)$ symmetry of the 
staggered Dirac operator (without the $U_A(1)$ symmetry) 
its smallest eigenvalues  thus have to approach their 
continuum limit as well.
%symmetry for staggered fermions in terms of the smallest Dirac eigenvalues.
%Close the continuum limit such that
%flavor symmetry is restored, the effective partition function
%is based on the symmetry group $U(4)\times U(4)$, and its supersymmetric
%extentition for the partially quenched partition function. For valence
%quark masses $m_v \ll m_c$, the partition function becomes a zero-dimensional
%group integral and the flavor symmetry group is encoded in the microscopic
%properties of the Dirac eigenvalues. If for example the distribution of the
%eigenvalues is according to the $U(1) \times U(1) $ symmetry we have not
%yet approached the continuum limit. 
%Let us make an estimate for the number
%of lattice points required for obtaining continuum Dirac spectra.
The staggered Dirac operator can be written as 
\be
D_{KS} = D_C + a^2 \Lambda^2 D_R,
\ee
where $D_C$ coincides with the continuum Dirac operator at low energies, 
$  a$ is the lattice spacing
and $\Lambda$ is a typical hadronic mass scale. The condition that
the Dirac spectrum of $D_{KS}$ approaches that of $D_C$ is 
(with $|| \cdot ||$ the norm of an operator)
\be
||a^2 \Lambda^2 D_R|| \ll \Delta \lambda a.
\label{condition}
\ee
With $\Sigma \sim \lambda^3$, 
the spacing of the eigenvalues near zero in lattice units is given by 
$\Delta \lambda a \sim 1/\rho(0) \sim 1/N\Lambda^{d-1} a^{d-1}$.
Since $||D_R|| \sim O(1)$, the
condition (\ref{condition}) can be rewritten as
\be
a^{d+1} \Lambda^{d+1} \ll \frac 1N \qquad {\rm or} \qquad
L \Lambda = N^{1/d} a \Lambda \ll N^{\frac 1d - \frac 1{d+1}}.
\ee
But we also require a sufficiently large lattice with $L\Lambda \gg 1$ 
resulting in 
\be
N^{\frac 1{d(d+1)}} \gg 1.
\ee
Our $naive$  estimate for the total number of lattice points for staggered
fermions to approach the continuum limit is given by $ N \approx (3^{d+1})^d$.
In two dimensions we need lattices of the order of $27^2$ to be reasonably
close to the continuum limit. This number is consistent with 
simulations of the Schwinger model with staggered fermions \cite{Hip}. 
In four dimensions the situation is much worse. 
According to the same estimate continuum physics is only seen on lattices
as large as $343^4$ which explains that todays staggered lattices show  
agreement with chRMT results in the sector of zero
topological charge \cite{Tilo,Heller,Damtopo,Phil}. 

\section{Conclusions}
We have argued that the the distribution of the smallest
eigenvalues of the Dirac operator is a signature for the 
pattern of chiral symmetry breaking of the QCD
partition function. Both the correlations of the smallest
eigenvalues and the slope of the Dirac spectrum from
have been obtained from a chiral Lagrangian. 
The intercept of the Dirac spectrum
determines the chiral condensate whereas its slope
fixes the pion decay constant. However, it takes very large staggered lattices
to reliably perform such analysis.

\vskip 0.5cm
\noindent {\bf Acknowledgments.} 
I gratefully acknowledge all my collaborators in this
project. D. Toublan is thanked for a critical reading of the manuscript.
This work was partially supported by the US DOE grant
DE-FG-88ER40388.


\begin{thebibliography}{9}
\bibitem{DeTar}C.~DeTar,
{\it Quark-gluon plasma in numerical simulations of QCD}, in {\it
Quark gluon plasma 2}, R. Hwa ed., World Scientific 1995.
\bibitem{LS}H.~Leutwyler and A.~Smilga, Phys. Rev. {\bf D46} (1992) 5607.
\bibitem{Hack}G. Hackenbroich and H.A. Weidenm\"uller, Phys. Rev. Lett.
{\bf 74} (1995) 4118.
\bibitem{ADMN}G. Akemann,
P. Damgaard, U. Magnea and S. Nishigaki, Nucl. Phys. {\bf B 487[FS]} (1997)
721.
\bibitem{Vplb}J.J.M. Verbaarschot, Phys. Lett. {\bf B368} (1996) 137.
\bibitem{Osbornprl}J.C. Osborn and J.J.M. Verbaarschot, Phys. Rev. Lett.
{\bf 81} (1998) 268.
\bibitem{OTV} J.C. Osborn, D. Toublan and J.J.M. Verbaarschot, Nuc. Phys.
{\bf B540} (1999) 317.
\bibitem{pqChPT} C. Bernard and M. Golterman, Phys. Rev. D49 (1994)
486; C. Bernard and M. Golterman, hep-lat/9311070; M. Golterman and
K.C. Leung, Phys. Rev. {\bf D57} (1998) 5703.
% {\it Partially quenched QCD and staggered fermions}.
\bibitem{DOTV} P.H. Damgaard, J.C. Osborn, D. Toublan and J.J.M. Verbaarschot,
Nucl. Phys. {\bf B547} (1999) 305.
\bibitem{Martin}M. Zirnbauer, J. Math. Phys. {\bf 37} (1996) 4986.
\bibitem{Altshuler} B.L. Altshuler, I.Kh. Zharekeshev, S.A. Kotochigova and
B.I. Shklovskii, Zh. Eksp. Teor. Fiz. {\bf 94} (1988) 343.
\bibitem{GL}J. Gasser and H.~Leutwyler, Phys. Lett. {\bf 188B} (1987) 477.
\bibitem{TV}D. Toublan and J.J.M. Verbaarschot, hep-th/9904199.
\bibitem{SVR}E.V. Shuryak and J.J.M. Verbaarschot,
Nucl. Phys. {\bf A560} (1993) 306; J.J.M. Verbaarschot,
Phys. Rev. Lett. {\bf 72} (1994) 2531.
\bibitem{Smilstern}A. Smilga and J. Stern, Phys. Lett. {\bf B318} (1993)
531.
\bibitem{Tilo} M.E. Berbenni-Bitsch, S. Meyer, A. Sch\"afer,
J.J.M. Verbaarschot and  T. Wettig, Phys. Rev. Lett. {\bf 80} (1998) 1146.
\bibitem{Hip} F. Farchioni, I. Hip, C.B. Lang, M. Wohlgenannt,
hep-lat/9812018.
\bibitem{Heller}R.G. Edwards, U.M. Heller and R. Narayanan,
hep-lat/9902021.
\bibitem{hiplat99}
F.~Farchioni, I.~Hip and C.B.~Lang,
%``Comparing lattice Dirac operators with random matrix theory,''
hep-lat/9907011.
\bibitem{Tiloval}
M.E.~Berbenni-Bitsch, M.~Gockeler, H.~Hehl, S.~Meyer, P.E.~Rakow,
A.~Schafer and T.~Wettig,
%``Random matrix theory, chiral perturbation theory, and lattice data,''
hep-lat/9907014.
\bibitem{Poulval}
P.H.~Damgaard, R.G.~Edwards, U.M.~Heller and R.~Narayanan,
%``Universal scaling of the chiral condensate in finite volume gauge
%                  theories,''
hep-lat/9907016.
\bibitem{Karlval}
P.~Hernandez, K.~Jansen and L.~Lellouch,
%``Finite size scaling of the quark condensate in quenched lattice QCD,''
hep-lat/9907022.
\bibitem{Damtopo}P.H.~Damgaard, U.M.~Heller, R.~Niclasen and
K.~Rummukainen,
%``Staggered fermions and gauge field topology,''
hep-lat/9907019.
\bibitem{Phil}P. de Forcrand, Aspen workshop on the Dirac Spectrum.
\end{thebibliography}
\end{document}